\input{aipcheck}

\documentclass[
    ,final            
  ]
  {aipproc}

\layoutstyle{6x9}

\usepackage{amsmath}
\begin{document}

\title{Holographic description of glueballs in a deformed AdS-dilaton background\footnote{Talk given at the International Workshop on
Quantum Chromodynamics: QCD@Work 2007, Martina Franca, Italy, 16-20
June 2007.}}


\author{Fr\'ed\'eric Jugeau}{
  address={INFN - Istituto Nazionale di Fisica Nucleare, Sezione di Bari, Italy (frederic.jugeau@ba.infn.it)}}

\begin{abstract}
We investigate the mass spectra of scalar and vector glueballs in
the so-called bottom-up approach of the AdS/QCD correspondence. The
holographic model of QCD includes a static dilaton background field.
We study the constraints on the masses coming from perturbing the
dilaton field and the geometry of the bulk.
\end{abstract}

\maketitle

\section{Introduction}
A breakthrough in the attempt to understand strongly coupled
Yang-Mills theories came with the AdS/CFT correspondence, proposed
by Maldacena \cite{Maldacena:1997re}, that conjectures a connection
between the large $N$ limit of a maximally $\mathcal{N}=4$
superconformal $SU(N)$ gauge theory defined in $d$ dimensions and
the supergravity limit of a superstring/M-theory living on a $d+1$
anti-de Sitter ($AdS$) space times a compact manifold
\cite{Witten:1998qj,Gubser:1998bc}. However, the application of this
conjecture to a theory such as QCD is not straightforward, being QCD
neither supersymmetric nor conformal so that its gravity dual theory
remains unknown. Witten proposed a procedure to extend Maldacena's
duality to such gauge theories \cite{Witten2:1998}: the conformal
invariance is broken by compactification, while supersymmetry is
broken by appropriate boundary conditions on the compactified
dimensions. The $AdS$ geometry of the dual theory is then deformed
into an $AdS$-Black-Hole geometry where the horizon plays the role
of an IR brane.

Adopting this (so-called up-down) approach, analyses of glueball
spectroscopy in 3 and 4 dimensions have been carried out, obtaining,
for instance, that the scalar glueball with $J^{PC}=0^{++}$
corresponds, in the supergravity side, to the massless dilaton field
propagating in the 10 dimension black-hole geometry \cite{Ooguri}.
Then, the glueball mass is computable by solving the dilaton wave
equation and gives results in reasonable agreement with the
available lattice data \cite{lattice}.

However, instead of trying to warp the original 11-dimensional
$AdS_7\times S^4$ geometry in order to obtain a 4-dimension gauge
theory with similarities with QCD, one could adopt a different
strategy investigating what features the dual theory should have in
order to reproduce known QCD properties. In this (so-called
bottom-up) approach, or AdS/QCD correspondence, one attempts to
construct a 5-dimensional holographic model able to reproduce the
main features of QCD. As pioneered by Polchinski and Strassler
\cite{polchinski}, it turned out to be possible to modify the
AdS/CFT duality, aimed at describing a confining gauge theory, by
considering a truncated $AdS_{5}$ holographic space-time on the
4-dimensional boundary of which QCD is defined. In this so-called IR
hard wall approximation, the typical size of this $AdS_{5}$ slice
stands for an IR cutoff associated to the QCD mass gap. The IR hard
wall model has been widely used in order to investigate, for
instance, light hadron spectrum and form factors \cite{radyu}.
Another holographic model of QCD has been proposed, which consists
in inserting a static dilaton field in the $AdS_{5}$ space-time.
This particular background allows one to recover the Regge behaviour
believed to be satisfied by mesons \cite{son2}, at odds with what
happens with the IR hard wall model (which rather appears to be dual
to a bag model of QCD) or starting from a general string theory and
attempting to deform it \cite{shifman}.

Even if somehow cumbersome shortcomings subsist when constructing
holographic models of QCD, namely, for instance, the stringy
corrections $O(1/N)$, the role of the remaining compact manifold
$S^{5}$ or the accurate range of the holographic coordinate that is
effectively dual to the QCD energy scale, there is the hope that
they do not spoil the main features of these dual models.

\section{The $5d$ holographic model dual to QCD}
Following \cite{son2}, we consider  a five dimensional conformally
flat spacetime (the bulk) described by the metric:
\begin{eqnarray}
g_{MN}=e^{2A(z)}\eta_{\,MN} &,& ds^2=e^{2A(z)}(\eta_{\,\mu\nu}dx^\mu
dx^\nu+dz^2)\;\;,
\end{eqnarray}
where $x^M=(x^\mu,z)$ and $\eta_{\,MN}=\hbox{diag}(-1,1,1,1,1)$.
$x^\mu$ ($\mu=0,\dots 3$) represent the usual space-time  (the
boundary) coordinates and $z$ is the fifth holographic coordinate
running from zero to infinity. The metric function $A(z)$ satisfies
the condition\footnote{In the following, we put the AdS radius $R$
to unity.}:
\begin{equation}\label{UV}
A(z)\underset{z\rightarrow0}{\rightarrow}\ln
\biggl(\frac{R}{z}\biggr)\;,
\end{equation}
to reproduce the $AdS_5$ metric close  to the UV brane $z \to 0$. In
the following, we will take the simplest choice compatible with the
constraints displayed in \cite{son2}: $A(z)=-\ln z$. Besides, we
consider a background dilaton field $\phi(z)=c^2 z^2$ which only
depends on the holographic coordinate $z$ and vanishes at the UV
brane. The large $z$ dependence of the dilaton is chosen to
reproduce the Regge behaviour of the low-lying mesons, and all the
masses will be given with respect to the scale parameter $c$.
Moreover, the introduction of this background dilaton allows one to
avoid ambiguities in the choice of the bulk field boundary
conditions at the IR wall.

We construct a $5d$ model that can be considered as a cut-off $AdS$
space: a smooth cut-off in the IR replaces the hard-wall IR cutoff
that would be obtained by allowing the holographic variable $z$ to
vary from zero to a maximum value
$z_m\simeq\frac{1}{\Lambda_{QCD}}$. To investigate the mass spectra
of the QCD  scalar and vector glueballs, we consider the two lowest
dimension operators with the corresponding quantum numbers and
defined in the field theory living on the $4d$ boundary
\cite{jugeau}:
\begin{equation}
\begin{cases}
O_{S}=Tr\,(F^2)\;,\\
O_{V}=Tr\,(F(DF)F)\;,\label{boundoper}
\end{cases}
\end{equation}
(with $D$ the covariant derivative) having conformal dimension
$\Delta=4$ and $\Delta=7$ respectively. The operator corresponding
to the vector glueball satisfies the Landau-Pomeranchuk-Yang
selection rule \cite{NSVZ}. According AdS/CFT correspondence, the
conformal dimension of a ($p$-form) operator on the boundary is
related to the $(AdS\;mass)^{2}$ of his dual field in the
 bulk as follows \cite{Witten:1998qj,Gubser:1998bc}:
\begin{equation}
(AdS\;mass)^2=(\Delta-p)(\Delta+p-4)\;.\label{m5}
\end{equation}
In the following, we assume that the mass $m_{5}^{2}$ of the bulk
fields is given by this expression.

A $5d$ massless  scalar field $X(x,z)$ can be constructed as the
correspondent of  $Tr\,(F^2)$, described by  the  action in the
gravitational background:

\begin{equation}
S=-\frac{1}{2}\int d^5x\,\sqrt{-g}\,e^{-\phi(z)}\,\,
g^{\,MN}(\partial_{M}X) (\partial_{N}X)\;,
\,\,\,\,\label{actionscalmass}
\end{equation}
with  $g=det(g_{MN})$. Scalar glueballs  are identified as the
normalizable modes of $X$ satisfying the equations of motion
obtained from \eqref{actionscalmass},  corresponding to a finite
action.

For the spin 1 glueball,   we introduce a 1-form $A_M$ described by
the action:
\begin{equation}
S=-\frac{1}{2}\int d^5x\,\sqrt{-g}\,e^{-\phi(z)}\,\biggl[
\frac{1}{2} g^{\,MN}g^{\,ST} F_{MS}
F_{NT}+m_5^2\,g^{\,ST}A_SA_T\biggr]\;,\label{actionvecmass}
\end{equation}
with $F_{MS}=\partial_{M} A_S - \partial_{S} A_M$ and  $m_5^2=24$,
and study its normalizable modes. Notice that the action
\eqref{actionvecmass}, with a different value of $m_5^2$,  describes
{\it a priori} fields that are dual to other operators in QCD,
namely those describing hybrid mesons with spin one,  which is  an
explicit example of different QCD operators having  similar bulk
fields as holographic correspondents.

\section{Scalar and vector glueball spectroscopy}

The field equations of motion obtained from the actions
(\ref{actionscalmass})-(\ref{actionvecmass}) can be reduced in the
form of a one dimensional Schr{\"o}dinger equation in the variable
$z$:

\begin{equation}
-\psi''+V(z)\psi=-q^2\psi\;,\label{geneqpot}
\end{equation}
involving the function $\psi(z)$ obtained
 applying a  Bogoliubov  transformation
$\psi(z)=e^{-B(z)/2}\tilde{Q}(q,z)$  to the Fourier transform
$\tilde{Q}$ of the field $Q$ ($Q=X, A_M$) with respect to the
boundary variables
 $x^\mu$. The function $B(z)$ is a combination of the dilaton
and the metric function: $B(z)=\phi(z)-a \, A(z)$,  with the
parameter $a$ given by:  $a=3$  and  $a=1$
 in  cases of $X$ and $A_M$ fields,  respectively.
The condition $q^2=-m^2$ identifies the mass of  the normalizable
modes of the two fields.

Eq. \eqref{geneqpot}  is a one dimensional Schr{\"o}dinger equation
where $V(z)$ plays  the role of a potential. It  reads as:

\begin{equation}
 V(z)=\frac{1}{4}\big(B'(z)\big)^2-\frac{1}{2}B''(z)+\frac{m_5^2}{z^2}=V_0(z)+\frac{m_5^2}{z^2}\;,\label{potscalar}
\end{equation}
with
\begin{equation}
V_0(z)= c^4 z^2+\frac{a^2+2a}{4 z^2}+c^2(a-1) \,\,
.\label{potentials}
\end{equation}
With this potential,  eq. \eqref{geneqpot} can be  analytically
solved. Regular solutions at $z\to 0$ and $z\to\infty$ correspond to
the spectrum:
\begin{equation}
  m_n^2=c^2\Big[4n+1+a+\sqrt{(a+1)^2+4m_5^2}\;\Big]\;,\label{specgenscal}
\end{equation}
with $n$ an integer (we identify it as a radial quantum number),
while the corresponding eigenfunctions read as:

\begin{equation}
\psi_n(z)=A_n\,e^{-c^2z^2/2}(cz)^{g(a,m_5^2)+1/2}\;{_1F_1}\left(-n,g(a,m_5^2)+1,c^2z^2\right)\;,\label{kummerscal}
\end{equation}
with  ${_1F_1}$ the Kummer confluent hypergeometric function,  $A_n$
a normalization factor and
$g(a,m_5^2)=\sqrt{\frac{(a+1)^2}{4}+m_5^2}$. From these relations,
we obtain the spectrum of  scalar and vector  glueballs,
respectively:
\begin{eqnarray}
m_n^2&=&c^2(4n+8)\;,  \label{specscal}\\
m_n^2&=&c^2(4n+12)\;.  \label{specvec}
\end{eqnarray}

A few remarks are in order in respect to the results derived in
\cite{son2}. First, both the spectra have the same dependence on the
radial quantum number $n$ as the mesons of spin $S$:
$m_n^2=c^2(4n+4S)$. This is a consequence of the large $z$ behaviour
chosen for the background dilaton. Second, both the lowest lying
gueballs are heavier than the $\rho$ mesons, the spectrum of which
reads: $m_n^2=c^2(4n+4)$. Finally, the vector glueball turns out to
be heavier than the scalar one.

More precisely, comparing our result to the computed $\rho$ mass, we
obtain for the lightest scalar   $(G_0)$  and vector $(G_1)$
glueballs
\begin{equation}
\frac{m^2_{G_{0}}}{m^2_\rho}=2 \hspace*{1cm};\hspace*{1cm}
\frac{m^2_{G_{1}}}{m^2_\rho}=3\;,\label{latticeres}
\end{equation}
which implies that these  glueballs are expected to be lighter than
as predicted by other QCD approaches \cite{lattice}. The result
$m^2_{G_1}-m^2_{G_0}=m^2_\rho$ predicts indeed a lightest vector
glueball with mass below 2 GeV.

It is interesting to investigate how it is possible to modify   the
$z$ dependence of the background dilaton field and of the metric
function $A$, and how the spectra change, an issue discussed in the
following section.
\section{Perturbed background}

There are other choices for the background dilaton $\phi$ and the
metric function $A$ which allow us to reproduce the Regge behaviour
of the low-lying mesons and to recover the $AdS_{5}$ metric close to
the UV brane when $z\rightarrow0$. As a matter of fact, it is
possible to add to the background fields terms of the type
$z^\alpha$ with $0\le\alpha<2$. Considering  the simplest case:
$\alpha=1$,  this can be done in two different ways. Either we
modify the dilaton field including a linear contribution which is
subleading in the IR regime $(z \to \infty)$ or we modify the metric
function which now acquires a linear term subleading in the  UV
regime $(z \to 0)$:
\begin{equation}
\left\{
\begin{array}{lll}
\phi(z)&=&c^2 z^2+\lambda c z\\
A(z)&=&-\ln z
\end{array}
\right.\hspace*{1cm},\hspace*{1cm}\left\{
\begin{array}{lll}
\phi(z)&=&c^2 z^2\\
A(z)&=&-\ln z- \lambda c z
\end{array}
\right.\;,\label{truc} \end{equation} with $\lambda$ a real
dimensionless parameter. The two choices produce different results.
Modifying the dilaton field, the potential \eqref{geneqpot} becomes:
\begin{equation}
V(z)=V_0(z) +\lambda V_1(z)+\frac{\lambda^2
c^2}{4}+\frac{m_5^2}{z^2}
f(z,\lambda)\hspace*{.5cm}\textrm{with}\hspace*{.5cm}\left\{
\begin{array}{lll}
V_1(z)&=&c\,(c^2 z+\frac{a}{2z})\\
f(z,\lambda)&=&1
\end{array}
\right.,\label{pert-pot}
\end{equation}
while modifying the metric in the IR, the potential term reads as:
\begin{equation}
V(z)=V_0(z) +\lambda \tilde V_1(z)+\frac{\lambda^2 c^2 a^2}{
4}+\frac{m_5^2}{z^2}
f(z,\lambda)\hspace*{.5cm}\textrm{with}\hspace*{.5cm}\left\{
\begin{array}{lll}
\tilde V_1(z)&=&a\,c \left(c^2 z+\frac{a}{2z} \right)\\
f(z,\lambda)&=&e^{-{2 \lambda cz }}
\end{array}
\right.,\label{v1-1}
\end{equation}
with $V_0(z)$  given in \eqref{potentials}. Considering
\eqref{truc}-\eqref{v1-1}, one sees that the mass term is the main
responsible of  the  difference between the scalar and vector cases
when the geometry is perturbed, while its effect turns out to be the
same when the background dilaton is modified. Eq. \eqref{geneqpot}
with the new potentials \eqref{pert-pot} and \eqref{v1-1}
 can be solved perturbatively and, for small values of the parameter
 $\lambda$, the spectra are modified:
\begin{equation}
  m_n^2=m_{n,(0)}^2+\lambda m_{n,(1)}^2 \,\,\, .\label{pert-mass}
\end{equation}
The detailed analysis can be found in \cite{jugeau}. Different
predictions at $O(\lambda)$ for the lowest-lying vector and scalar
glueball mass difference are then obtained, modifying either the
dilaton or the geometry:
\begin{eqnarray}
m_{G_1}^2-m_{G_0}^2&=&c^2\big(4-\frac{3\sqrt\pi}{128}
\lambda\big)\hspace*{2,2cm}\textrm{(modifying the dilaton)}\;\;,\\
m_{G_1}^2-m_{G_0}^2&=&c^2\big(4-\frac{1899  {\sqrt \pi}}{128}
\lambda\big) \hspace*{1cm}\textrm{(modifying the metric
function)}\;\;.\label{res}
\end{eqnarray}
Therefore, the mass splitting between vector and scalar glueballs
increases if $\lambda$ is negative, and the maximum effect is
produced, for the same value of  $\lambda$, when the metric function
is perturbed. This can be considered as an indication on the type of
constraints the background fields in the bulk must satisfy.

\section{Conclusions}
We have discussed how  the QCD holographic model proposed in
\cite{son2}, with the hard IR wall replaced by a background dilaton
field, allows one to predict  the  light glueball spectra
\cite{jugeau}. Vector glueballs turn out to be heavier than the
scalar ones, and the dependence of their masses on the radial
quantum number is the same as obtained for  $\rho$ and higher spin
mesons. Combining the calculations of the glueball and  $\rho$
masses in the same holographic model, the glueballs  turn out to be
lighter than predicted in other approaches \cite{lattice}.

We have also investigated how the masses change as a consequence of
perturbing  the dilaton in the UV or the bulk geometry in the IR,
finding that constraints in the bottom-up approach can be found if
information on the spectra from other approaches is considered. Such
constraints should be taken into account in the  attempt to
construct the QCD gravitational dual.

\begin{theacknowledgments}
I am very indebted to my collaborators at the INFN sezione di Bari
who make my stay so pleasant and fruitful. I am also grateful to the
organizers who gave me the opportunity to present my research
activities during this workshop. This work was supported in part by
the EU Contract No. MRTN-CT-2006-035482, "FLAVIAnet".
\end{theacknowledgments}



\bibliographystyle{aipproc}   


\begin{thebibliography}{99}
\bibitem{Maldacena:1997re}
  J.~M.~Maldacena, \emph{The large N limit of superconformal field theories and
supergravity},
  Adv.\ Theor.\ Math.\ Phys.\  \textbf{2} (1998) 231
  [Int.\ J.\ Theor.\ Phys.\  \textbf{38} (1999) 1113] (hep-th/9711200).

\bibitem{Witten:1998qj}
  E.~Witten,
  \emph{Anti-de Sitter space and holography},
  Adv.\ Theor.\ Math.\ Phys.\  {\bf 2} (1998) 253.

\bibitem{Gubser:1998bc}
  S.~S.~Gubser, I.~R.~Klebanov and A.~M.~Polyakov,
  \emph{Gauge theory correlators from non-critical string theory},
  Phys.\ Lett.\ B {\bf 428} (1998) 105
  (hep-th/9802109).

\bibitem{Witten2:1998}
E. Witten, \emph{Anti-de Sitter space, thermal phase transition, and
confinement in gauge theories}, Adv.\ Theor.\ Math.\ Phys.2, 505,
1998 (hep-th/9803131).

\bibitem{Ooguri}
C. Csaki, H. Ooguri, Y. Oz, J. Terning, \emph{Glueball mass spectrum
from supergravity}, JHEP 9901 (1999) 017 (hep-th/9806021).

\bibitem{lattice}
C.~J.~Morningstar and M.~J.~Peardon, \emph{The glueball spectrum
from an anisotropic lattice study}, Phys.\ Rev.\ D {\bf 60} (1999)
034509 (hep-lat/9901004); B.~Lucini and M.~Teper, \emph{SU(N) gauge
theories in four dimensions: Exploring the approach to infinity},
JHEP {\bf 0106} (2001) 050 (hep-lat/0103027).

\bibitem{polchinski}
J. Polchinski, M. J. Strassler, \emph{Hard scattering and
gauge/string duality}, Phys.\ Rev.\ Lett. {\bf 88} (2002) 031601
(hep-th/0109174).

\bibitem{radyu}
H. R. Grigoryan and A. V. Radyushkin, \emph{Form Factors and Wave
Functions of Vector Mesons in Holographic QCD} (hep-ph/0703069).

\bibitem{son2}
  A.~Karch, E.~Katz, D.~T.~Son and M.~A.~Stephanov, \emph{Linear confinement and
  AdS/QCD}, Phys.\ Rev.\ D {\bf 74} (2006) 015005 (hep-ph/0602229).

\bibitem{shifman}
E. Schreiber, \emph{Excited mesons and quantization of string
endpoints} (hep-th/0403226); M. Shifman, \emph{Highly excited
hadrons in QCD and beyond}, Frascati 2005, Quark-hadron duality and
the transition to pQCD p.171-191 (hep-ph/0507246).

\bibitem{jugeau}
P. Colangelo, F. De Fazio, F. Jugeau and S. Nicotri, \emph{On the
light glueball spectrum in a holographic description of QCD}, Phys.
Lett. {\bf B}652 (2007) 73-78 (hep-ph/0703316).


\bibitem{NSVZ}
V. A. Novikov, M. A. Shifman, A. I. Vainshtein and V. I. Zakharov,
\emph{Are All Hadrons Alike?}, Nucl. Phys. B {\bf 191} (1981) 301.




\end{thebibliography}

\IfFileExists{\jobname.bbl}{}
 {\typeout{}
  \typeout{******************************************}
  \typeout{** Please run "bibtex \jobname" to optain}
  \typeout{** the bibliography and then re-run LaTeX}
  \typeout{** twice to fix the references!}
  \typeout{******************************************}
  \typeout{}
 }


\end{document}